 \documentclass[smallabstract,smallcaptions]{dccpaper}

\usepackage{epsfig}
\usepackage{citesort}
\usepackage{amsmath}
\usepackage{amssymb}
\usepackage{color}
\usepackage{url}

\newlength{\figurewidth}
\newlength{\smallfigurewidth}

\begin{document}

\title
{\large
\textbf{Conditional Neural Video Coding with Spatial-Temporal Super-Resolution
}
}

\author{%
Henan Wang, Xiaohan Pan, Runsen Feng, Zongyu Guo, and Zhibo Chen\\[0.5em]
{\small\begin{minipage}{\linewidth}\begin{center}
\begin{tabular}{c}
University of Science and Technology of China \\
Hefei, Anhui, China \\
\url{chenzhibo@ustc.edu.cn}
\end{tabular}
\end{center}\end{minipage}}
}

\maketitle
\thispagestyle{empty}

\begin{abstract}
This document is an expanded version of a one-page abstract originally presented at the 2024 Data Compression Conference. It describes our proposed method for the video track of the Challenge on Learned Image Compression (CLIC) 2024. Our scheme follows the typical hybrid coding framework with some novel techniques. Firstly, we adopt Spynet network to produce accurate motion vectors for motion estimation. Secondly, we introduce the context mining scheme with conditional frame coding to fully exploit the spatial-temporal information. As for the low target bitrates given by CLIC, we integrate spatial-temporal super-resolution modules to improve rate-distortion performance. Our team name is IMCLVC.

\end{abstract}

\Section{Introduction}

In the video track of CLIC2024, participants are required to submit a codec to compress a set of 10-second sequences with diverse contents and frame rates. The winners will be chosen based on human perceptual visual quality. To meet the requirements, we propose an efficient neural video  codec optimized for MS-SSIM. In this paper, we briefly describe the design and training details.

\Section{Related Works}

In recent years, the field of learning-based video compression has witnessed remarkable advancements. These methods predominantly follow the conventional hybrid coding architecture while integrating neural networks to enhance various aspects such as motion compensation and context analysis. DVC\cite{lu2019dvc} is the first end-to-end video compression framework. The model is designed by replacing every component in a traditional video compression framework with neural networks. A notable feature of DVC is its incorporation of optical flow networks for motion estimation, setting a foundational architecture for subsequent variations. Building on this architecture, several variations have emerged, each addressing specific challenges in video compression. SSF\cite{agustsson2020scale} introduces a scale parameter to the optical flow, enhancing the network's ability to model uncertainty. This adaptation proves particularly effective in scenarios involving disocclusions or rapid motion. FVC\cite{hu2021fvc} conducts key video compression operations, including motion estimation, motion compression, motion compensation, and residual compression, within the feature space. ELF-VC\cite{rippel2021elf} offers an innovative flexible-rate framework, enabling adjustable rate control within a single model. Incorporating reference modes from traditional video compression, LHBDC\cite{yilmaz2021end} integrates bi-directional prediction as well as hierarchical coding architecture into learning-based video compression frameworks. C2F\cite{hu2022coarse} presents a coarse-to-fine framework accompanied by hyperprior-guided mode prediction methods. This approach significantly enhances motion compensation and optimizes coding modes. ENVC\cite{guo2023learning} designs a reference feature pyramid as prediction sources, aiming to refine the precision of prediction. Additionally, it introduces a weighted prediction mechanism focused on the reconstruction of fine details.

The recent shift in video compression strategies is marked by the growing interest in a new paradigm known as conditional coding, first introduced in DCVC\cite{li2021deep}. This innovative approach harnesses contextual information derived from motion compensation to enhance the encoding and decoding processes. It represents a shift from explicit residual coding, where the focus is on encoding the differences (residuals) between frames, to a more nuanced strategy known as implicit conditional coding. Further advancements are seen in DCVC-TCM\cite{sheng2022temporal}, which advances this concept by enabling the learning of multi-scale temporal contexts for more effective information transmission. Subsequent works\cite{ho2022canf,qi2023motion,li2023neural} continue to explore various model architecture designs, further enriching the potential and scope of conditional coding in video compression.

\Section{Proposed Methods}

\SubSection{Spatial-temporal Super-Resolution}

\begin{figure}[t]
\begin{center}
\includegraphics[width=0.75\textwidth]{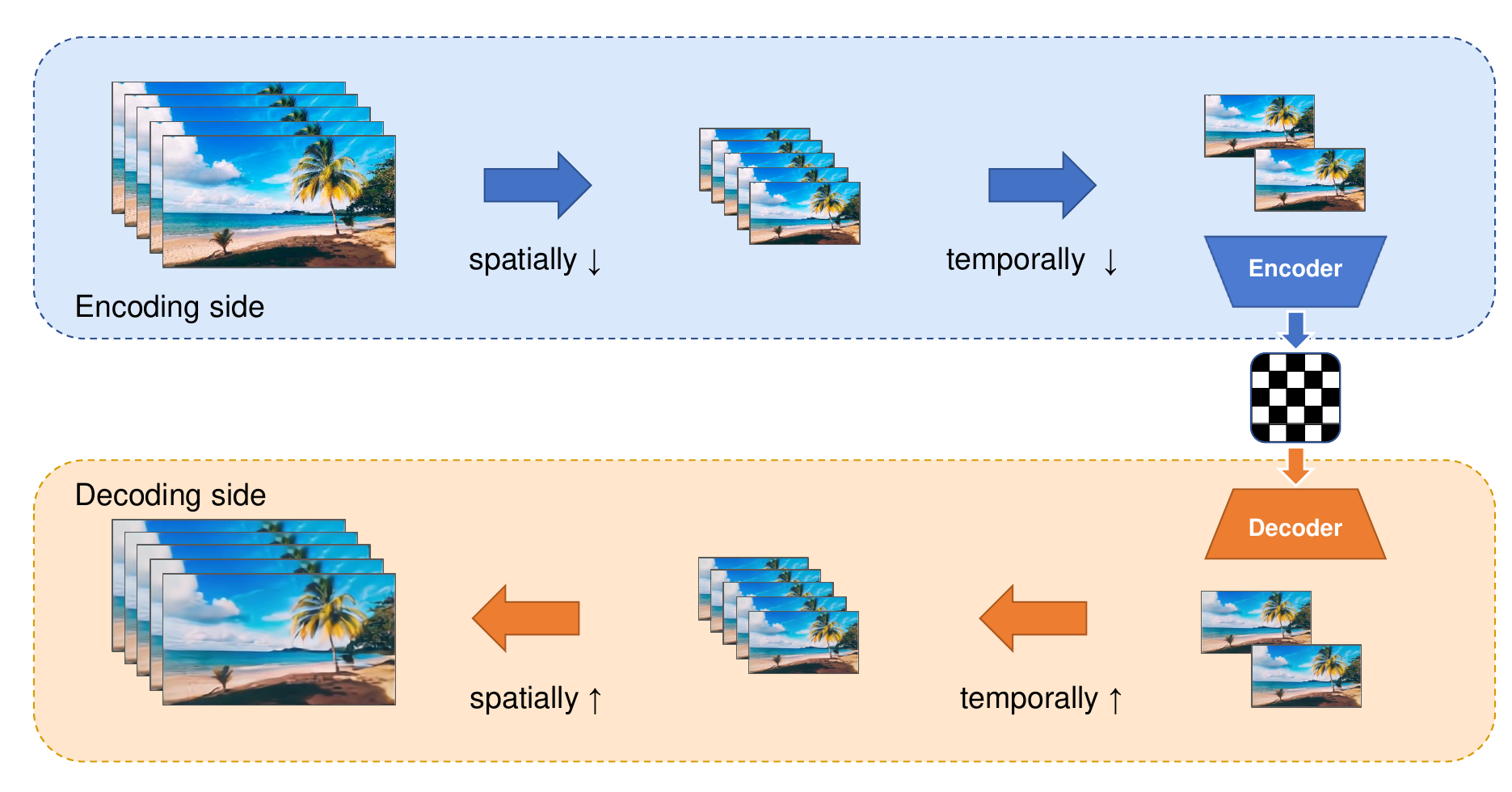}
\end{center}
\caption{\label{fig:sr}%
Proposed spatial-temporal super resolution scheme. Videos are downsampled spatially and temporally before being coded, while upsampled to original resolution at the decoder side.}
\end{figure}

To enhance the rate-distortion (RD) performance, particularly at lower bitrates, a novel approach combining downsampling and super-resolution techniques has been implemented, as shown in Fig \ref{fig:sr}.

For the ultra-low bitrate scenario of 0.05 mbps, the input video undergoes a downsampling process in both spatial and temporal dimensions, specifically by a factor of 4. This drastic downsampling results in a substantially smaller video file, which is then encoded and transmitted efficiently.

At the decoder side, we first decode the downsampled video. Following this, an advanced frame interpolation technique EMA-VFI\cite{zhang2023extracting} is applied to reconstruct the omitted frames, compensating for the temporal information lost during downsampling. Subsequently, to address the spatial downsampling, bicubic interpolation is employed. 

In the higher bitrate of 0.5 mbps, only temporal downsampling and interpolation are utilized, leaving the spatial resolution intact. By doing so, the texture and detail of the original video are better preserved.

\begin{figure}[t]
\begin{center}
\includegraphics[width=0.75\textwidth]{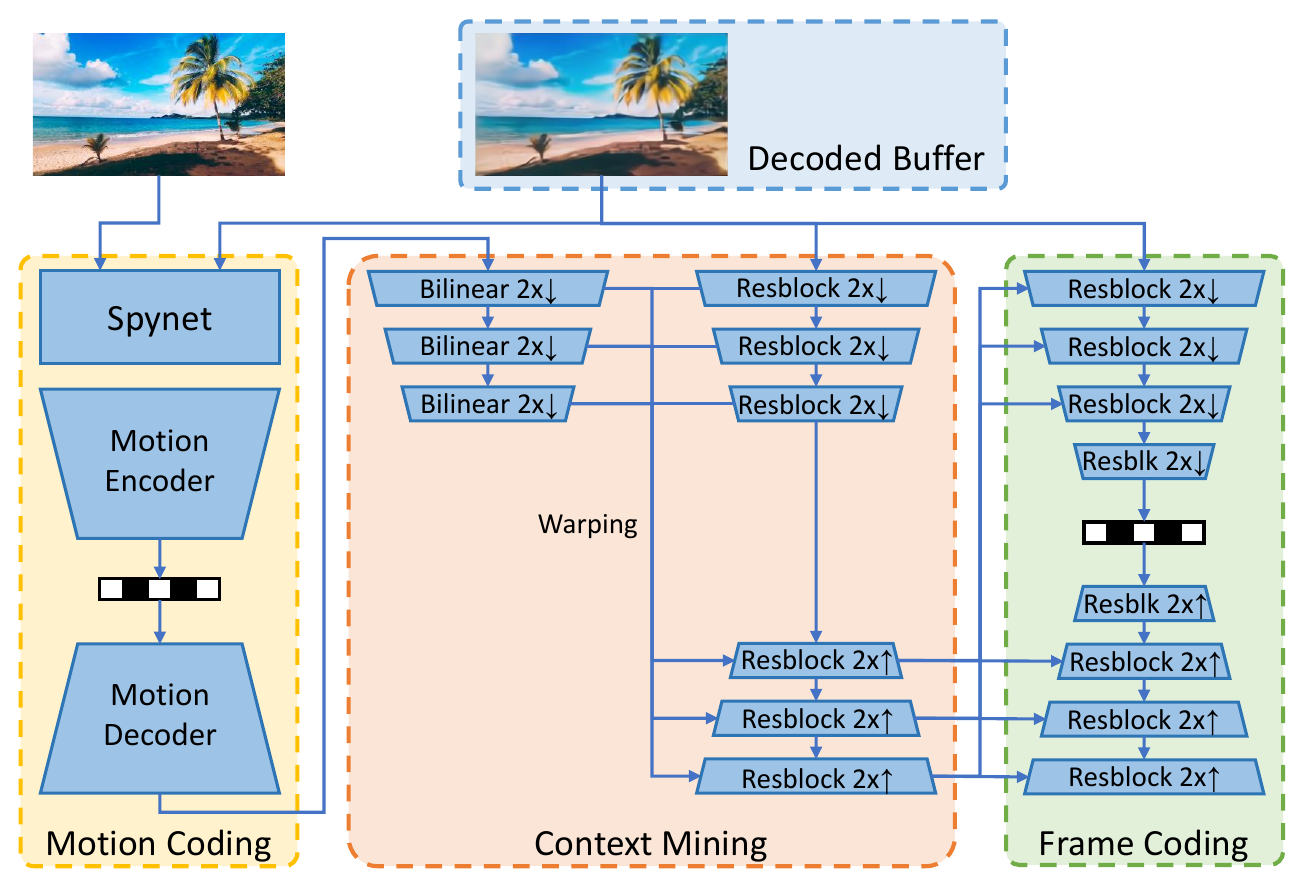}
\end{center}
\caption{\label{fig:framework2}%
Our P-frame model architecture. The model is composed of three parts: motion coding, context mining and frame coding. For motion estimation, we adopt the Spynet \cite{ranjan2017optical}. For temporal context mining, we adopt the TCM module in DCVC-TCM \cite{sheng2022temporal}.}
\end{figure}

\SubSection{Codec Architecture}

The proposed end-to-end optimized video codec follows the typical hybrid coding framework with some new techniques. In particular, we adopt the conditional coding framework. As shown in Fig \ref{fig:framework2}, the main body can be divided into three parts: motion coding, context mining and frame coding. 

1) Motion Coding: The optical flow produced by the pretrained Spynet \cite{ranjan2017optical} is taken as the estimated motion vector, which is then encoded and decoded by an hyperprior-based autoencoder.

2) Context Mining: The rich spatial-temporal context information provided by the decoded frame and motion vector should be fully extracted. Basically we adopt the temporal context mining (TCM) module from DCVC-TCM \cite{sheng2022temporal}. The TCM module uses multi-scale warping to generate multi-scale temporal contexts and capture spatial-temporal non-uniform motion and texture.

3) Frame Coding: The spatial-temporal context feature generated by the context mining module is introduced as the conditional information for frame coding. Specially, the contextual encoder and contextual decoder take the conditional features in at different scale to fully utilize the conditional information. Our context model is developed based on the unevenly grouped channel-wise context model in \cite{he2022elic}. This helps our models strike a balance between complexity and performance.

\SubSection{Training Strategy}

The training procedure of our codec can be divided into 2 stages. At the first stage, the loss calculation and model optimization is carried out per frame. The I-frame model is trained firstly and freezed when we are training the P-frame model. For P-frame training, we also learn from the multi-stage training strategy proposed in DCVC-TCM \cite{sheng2022temporal}, where the motion codec is trained seperately from the frame codec. With the help of this training strategy, we pre-train the model at high rates and then finetune the model to the target rate at the next stage.


The next stage is multi-frame co-optimization. Here we jointly optimize the I-frame model and P-frame model. In our low-delay setting, the latter P-frame uses previously reconstructed P-frame or I-frame as reference. Since the reference frame is reconstructed with some loss at the decoder side, the error will accumulate as the number of picture order count (POC) increases. To mitigate the error propagation issue, we jointly optimize the I-frame model and P-frame model, making the model aware of error propagation. Furthermore, we adjust the training GoP size as training proceeds, enabling the model to be better equipped for handling larger GoP sizes during testing. We first set the training GoP size to 3 and later to 5. The overall Rate-Distortion (RD) optimization target function is shown in Eq. \ref{eq:loss} 
\begin{equation}
    \label{eq:loss}
    \mathcal{L}=\mathcal{R}_I+\lambda\cdot\mathcal{D}_I+\sum_{t=1}^{T-1}\left(\mathcal{R}_t+\lambda\cdot\mathcal{D}_t\right),
\end{equation}
where $\mathcal{R}_I$-$\mathcal{D}_I$ and $\mathcal{R}_t$-$\mathcal{D}_t$ represent the rate-distortion of the I-frame and the $t$-th P-frame, $T$ represents the training GoP size. Since perceptual quality is evaluated in CLIC, we choose MS-SSIM as our distortion objective.

\Section{References}
\bibliographystyle{IEEEbib}
\bibliography{refs}

\end{document}